\documentclass[twocolumn]{svjour3}          % twocolumn
\smartqed  % flush right qed marks, e.g. at end of proof
\usepackage{graphicx}

\begin{document}

\title{Particle acceleration by shocks in supernova remnants 
}

\author{AR Bell 
%\\
%Clarendon Laboratory, University of Oxford, Oxford OX1 3PU, UK \\
%email: t.bell1@physics.ox.ac.uk
}


\institute {Clarendon Laboratory, University of Oxford, Oxford OX1 3PU, UK \\
              \email{t.bell1@physics.ox.ac.uk}           %  \\
}
\date{Received: date / Accepted: date}

\maketitle

\begin{abstract}
Particle acceleration occurs on a range of scales 
from AU in the heliosphere to Mpc in clusters of galaxies
and to energies ranging from MeV to EeV. 
A number of acceleration processes have been proposed, 
but diffusive shock acceleration (DSA) is widely invoked as the predominant mechanism.  
DSA operates on all these scales and probably to the highest energies.  
DSA is simple, robust and predicts a universal spectrum. 
However there are still many unknowns regarding particle acceleration. 
This paper focuses on the particular question of
whether supernova remnants (SNR) can produce the
Galactic CR spectrum up to the knee at a few PeV.
The answer depends in large part 
on the detailed physics of diffusive shock acceleration.  

\end{abstract}

%===============================================================================================

\section{Introduction}
\label{intro}
Energetic particles are prevalent throughout the universe, ranging from MeV particles in the heliosphere to TeV particles in
supernova remnants to EeV particles that almost certainly originate beyond our own Galaxy.
At low energies within the heliosphere they are detected by satellites,
while at high energies they arrive at the Earth as cosmic rays (CR).
CR in distant objects are detected from radio to $\gamma$-ray wavebands
by means of electron synchrotron and inverse Compton emission 
or $\gamma$-ray emission from pion-producing collisions between cosmic rays and
background protons.
The  presence of CR in SNR with energies of 100s of TeV is
proved by the detection at the Earth of $\gamma$-rays with energies up to nearly 100TeV
arriving from the direction of the source (Aharonian et al 2007, 2011).
Particle acceleration is known to be efficient because the Galactic cosmic ray energy density
can only be generated if a significant fraction of the energy of supernova explosions
is given to energetic particles.
Blandford and Eichler (1987) estimate
an efficiency of a few percent.
Initial explanations of CR acceleration centred on the second order Fermi process (Fermi 1949)
by which CR gain energy from elastic collisions with magnetic field structures in motion
in the interstellar medium (ISM). 
If the typical velocity of a magnetic structure is $u$, the excess probability of a CR encountering
it head-on rather than tail-on is proportional to $u/c$ with a fractional increase in CR energy 
of order $u/c$ on a head-on encounter, so the process is second order in $u/c$.
The second order process probably contributes to particle acceleration, but
it was realised in the late 1970s that a more rapid first order process,
known as diffusive shock acceleration (DSA), 
occurs wherever there are strong releases of energy
that produce blast waves (Krymsky 1977, Axford et al 1977, Bell 1978, Blandford \& Ostriker 1978).
CR diffusion back and forth across a strong shock produces a fractional energy gain
of order $u/c$ on each crossing where $u$ is the shock velocity.
On average each CR crosses the shock $c/u$ times before escaping downstream of the shock,
with a probability $ (1-u/c)^m$ of escaping after at least $m$ shock crossings.
A simple derivation shows that a $E^{-2}$ differential energy spectrum
arises naturally from DSA at a strong shock.
Its advantages as a theoretical explanation of particle acceleration
are that a large part of a supernova energy release is processed through the expanding shock front,
the energy spectrum matches observation reasonably well,
and high acceleration efficiency appears likely.
The conclusion was soon drawn that DSA at supernova shocks is probably
the main source of Galactic CR.
The highest energy (EeV) CR may possibly be produced 
in AGN, radio galaxies or GRB
where much of the available energy is processed through high velocity shocks
or shocks with a large spatial extent.

DSA offers a generic explanation of astrophysical particle acceleration,
but the explanation has not been, and still is not, completely straightforward.
In two crucial papers,
Lagage \& Cesarsky (1983a,b)
reasoned that DSA at SNR shocks appeared incapable of accelerating protons
to PeV energies.
Since the Galactic CR energy spectrum extends
from GeV to a few PeV as a nearly straight power law, 
it seems necessary that the same 
mechanism accelerating CR to TeV should also be accelerating 
CR from TeV to PeV.
Because the spectrum steepens rather than flattens in a `knee' at a few PeV 
the spectrum beyond the knee cannot easily be explained
as the coincidental meeting of different populations above and below the knee.
So the same process is probably responsible for CR acceleration beyond the knee,
although the problem is alleviated if CR beyond the knee are heavy nuclei rather than protons.

This review is principally concerned with 
the particular question of whether supernova remnants (SNR)
accelerate CR to the knee in the Galactic energy spectrum at a few PeV.
More broadly based reviews of diffusive shock acceleration 
or with a different focus can be found in Drury (1983), 
Blandford \& Eichler (1987), Jones \& Ellison (1991),
Malkov \& Drury (2001) and Bell (2013).

%=========== SECTION 2 ====================================================================================

\section{The maximum CR energy}
In DSA at a strong shock propagating at velocity $u$
the CR energy increases on average by a fraction $u/c$
each time the CR cycles from upstream to downstream and back to upstream.
Consequently a CR must cross the shock typically $(c/u)\log (E_2/E_1)$ times for
acceleration from energy $E_1$ to $E_2$.
Krymsky (1979) derived a time $\lambda/u_d$,
where $\lambda$ is the CR mean free path and $u_d$ is the downstream velocity relative to the shock,
for CR to cycle between upstream and downstream.
Allowing for different conditions upstream
and downstream, the complete mean cycle time
between upstream and downstream is
$\tau_{cycle}=4[D_u/u+D_d/u_d]/c$ (Lagage \& Cesarsky (1983a,b))
where $D_u$ and $D_d$ are the upstream and downstream CR diffusion coefficients
respectively, and $u_d=u/4$ for a strong shock.
The ratio of $\tau_{cycle}$ to the age of the SNR indicates that
CR reach a maximum energy
of typically  $\sim 10^{13}{\rm eV}$, which falls 100 times short of the energy of the knee.
The Lagage \& Cesarsky limit can be derived as follows.

Balance between diffusion away from the shock upstream and advection back to the shock at velocity
$u$ creates an exponential CR precursor upstream of the shock with a scaleheight $L=D_u/u$.
The number of CR in the precursor per unit shock area is $nD_u/u$
where $n$ is the CR number density at the shock.
CR cross the shock at a rate $nc/4$ so the average time a CR spends upstream between crossings is
$4D_u/cu$.  The time spent downstream is similarly $4D_d/cu_d$ giving the total cycle time.
At each cycle the CR energy increases by a fraction $u/c$ on average
so the characteristic time for energy gain is $\tau _{gain}=(c/u)\tau _{cycle}= 4(D_u+4D_d)/u^2$.

Blasi \& Amato (2012) find best fits for the Galactic diffusion coefficient
of the form $D = D_{28} 10^{28} (E/3GeV)^\delta$
$ {\rm cm}^2{\rm s}^{-1}$
with $D_{28}=1.33 H_{kpc}$ for $\delta=1/3$ or $D_{28}=0.55 H_{kpc}$ for $\delta=0.6$, 
where $H_{kpc}$ is the height of the Galactic halo in kpc.
If these diffusion coefficients were to apply to diffusive shock acceleration in SNR
expanding at 5000 ${\rm km\ s}^{-1}$
the acceleration time at GeV energies would be $\sim 10^4\ {\rm years}$.
It would take an impossibly long time for
CR to reach PeV energies, between $10^6$ and $10^8\ {\rm years}$, 
by which time the shock would have slowed substantially 
($u\ll 5000 {\rm km\ s}^{-1}$)
or dissolved into the interstellar medium.
The long estimated acceleration time reflects the large mean free path for CR scattering in
the interstellar medium.
Fortunately, the large currents of streaming CR close to the SNR
excite plasma instabilities that 
drive fluctations in the magnetic field and increase CR scattering.
Two especially effective CR-driven plasma instabilities 
are a resonant instability (Lerche 1967, 
Kulsrud \& Pearce 1969, Wentzel 1974, Skilling 1975a,b,c
and Bell 1978), in which the CR resonantly excite Alfven waves,
and a non-resonant instability (Bell 2004),
in which a purely growing mode is strongly driven under 
conditions found in young SNR.
Throughout this paper the first of these two instabilities
will be referred to as the `Alfven instability' because it drives
Alfven waves.
The second instability will be referred to as the `non-resonant hybrid (NRH)
instability' because the mode is not spatially resonant with the CR Larmor radius
and because it is hybrid in the sense that kinetic CR interact with a fluid background plasma.
These instabilities will be discussed below.
Other instabilities such as the Drury instability (Drury \& Falle 1986),
the firehose instability, and other long wavelength instabilities
(Bykov et al 2011, Reville \& Bell 2012, 2013, Schure \& Bell 2011)
probably also play a role in perturbing the upstream magnetic field
and reducing the diffusion coefficient.
Additionally, magnetic field amplification may take place downstream of the shock due to 
vorticity induced by density
inhomogeneities in the plasma overtaken by the shock (Giacalone \& Jokipii (2007); 
Zirakshvili \& Ptuskin (2008) )
but amplification may occur too far downstream to contribute to CR acceleration at the shock
(Guo et al 2012).

Plasma instabilities can generate magnetic structures that scatter CR,
but the mean free path for scattering in a magnetic field cannot be smaller than the Larmor
radius $r_g=p/eB$ where $p$ is the CR momentum
and $B$ is the magnetic field upstream of the shock in Tesla.
The upstream diffusion coefficient $D_u$ is consequently at least $r_gc/3=E/3B$ where $E$ is the
energy in eV.
The perpendicular component of the magnetic field is compressed by a factor of four
at the shock so a reasonable assumption is that 
$D_d\sim D_u/4$ within a numerical factor of order one.
If $D_d = D_u/4$
the downstream cycle time 
$4D_d/c(u/4)$ is equal to the upstream cycle time 
$4D_u/cu$
and the total cycle time is $\tau _{cycle}=8E/3Bu$.
Under this assumption,
the characteristic time for energy gain is $\tau _{gain}= 8E/3Bu^2$.
Setting the acceleration time equal to the SNR expansion time $R/u$
where $R$ is the SNR radius gives a maximum energy
$E_{max}=3uBR/8\ $eV,
which is equivalent to
$$
E_{max}=1.2\times 10^{13}u_7 B_{\mu G} R_{pc}\ {\rm eV}
\eqno{(1)}
$$ 
where $B_{\mu G}$ is the magnetic field in microGauss,
$R_{pc}$ is the SNR radius in parsec,
and $u_7$ is the shock velocity in $10^7{\rm m s}^{-1}$.
For Cas A, $u_7=0.6$ and $R_{pc}=1.7$, giving $E_{max}\approx 60 {\rm TeV}$
for a typical interstellar magnetic field of $5\mu {\rm G}$.
Similarly for Tycho's SNR, $E_{max}\approx 130 {\rm TeV}$
assuming $u_7=0.5$ and $R_{pc}=4.3$ (following Williams et al 2013).
$E_{max}$ falls far short of the knee for both these SNR,
thus illustrating the challenge posted by Lagage \& Cesarsky (1983a,b)
to DSA as an explanation of Galactic CR acceleration.
These estimates of $E_{max}$ are further reduced by a factor of three
if the diffusion coefficient is set to $r_gc$ instead of 
the probably extreme $r_gc/3$ assumed above.

 In the Sedov phase of SNR expansion the radius increases but the shock velocity decreases
 such that
 $\rho u^2R^3=0.35 \epsilon $, where $\epsilon$ is the energy of a Sedov blast wave
 expanding into a medium of density $\rho$.
 This implies that the maximum CR energy decreases during the Sedov phase:
 $E_{max}=3\times 10^{13}B_{\mu G}\epsilon_{44}^{1/2}n_e^{-1/2} R_{pc}^{-1/2} {\rm eV}$
 where $\epsilon_{44}$ is in units of $10^{44}{\rm J}$ and $n_e$ is the ambient density
 in ${\rm cm}^{-3}$.
 CR may be accelerated to higher energy
 if they are able to stay with the shock throughout the Sedov phase and
 gain energy continually, but such high energy CR are few in number because
 the probability of staying with the shock for a long time is very small, 
 and their final energy is not much increased beyond their energy at the start
 of the Sedov phase.

 The difficulties posed by Lagage \& Cesarsky are clearly seen to be stringent
 when it is recognised that the Bohm diffusion coefficient $D=r_gc/3$
 is a lower limit for CR diffusion along magnetic field lines.
Two conditions must hold for Bohm diffusion to apply.
Firstly, fluctuations in the magnetic field $\delta B$ must be at least comparable
in magnitude with any mean field, $\delta B/B \ge 1$.
Secondly, the scalelength of the fluctuations must match the CR Larmor radius.
CR barely notice fluctuations on a scale much smaller scale than
the Larmor radius.
On the other hand, 
if the scalelength is much larger than the Larmor radius, the CR will see them as a mean field and
spiral freely along the total field without scattering.
However a mixture of theoretical (Reville et al 2008)
and observational studies (Stage et al 2006, Uchiyama et al 2007)
indicate that
the assumption of Bohm diffusion is not unreasonable.

%================ SECTION 3 ===================================================================
\section{Oblique \& perpendicular shocks}

According to Lagage \& Cesarsky the maximum CR energy is limited by
the ratio of the cycle time $\tau_{cycle}$ to the age of the SNR.
Jokipii (1987) showed that the cycle time can be much shorter for `perpendicular shocks'
where the upstream magnetic field is perpendicular to the shock normal.
In the intermediate case of an oblique shock with an angle $\theta$ between the magnetic field
and the shock normal
a CR must propagate at a projected velocity $u/\cos \theta$ along a field line 
if it is to outrun the shock.
Consequently, CR are overtaken by the shock in time $4D_u/(u/\cos \theta))$, 
the acceleration rate is increased by $1/\cos \theta$, and CR reach an energy which is
higher by $1/\cos \theta$ in the same time $R/u$.
This is still not sufficient to account for acceleration to a PeV unless $\theta$
is nearly $\pi/2$.

Acceleration can be most rapid when the shock
is perpendicular ($\theta = \pi/2$).
In this case CR cannot outrun the shock by propagating along a field line
and CR must diffuse across magnetic field lines to penetrate into the upstream plasma.
This is possible if CR are scattered by fluctuations in the magnetic field.
In the standard theory for charged particle diffusion across a magnetic field,
the cross-field diffusion coefficient is $D_\perp = D_{||}/(\omega _g \tau _{scat})^2$
where $\omega _g$ is the Larmor frequency, $\tau _{scat}$ is the scattering (collision) time
for the CR in the perturbed magnetic field, and $D_{||}$ is the diffusion coefficient along
the magnetic field.
The CR acceleration rate is thereby increased by a factor $(\omega _g \tau _{scat})^2$
although acceleration ceases if $\omega _g \tau _{scat} > c/u$ in which case
scattering is not sufficiently strong for CR to penetrate into the upstream plasma
by one Larmor radius.
Under optimal conditions and a suitable value of $\omega _g \tau _{scat}$
the maximum CR energy as defined by the acceleration time can be greatly increased
(Jokipii 1987, Ellison et al 1995, Meli \& Biermann 2006).

Unfortunately a different constraint intervenes to limit the maximum CR energy
at a perpendicular shock.
CR gain energy by drifting along the shock surface
under the influence of the magnetic field discontinuity between the upstream
magnetic field and the compressed downstream magnetic field.
The source of the energy gain is the ordered $-{\bf u}\times {\bf B}$ electric field
necessary for the thermal plasma to cross the magnetic field at velocity $\bf u$
in the rest frame of the shock.
The maximum distance the CR can drift on the shock surface is the shock radius $R$ yielding a maximum
CR energy $uBR$ which is very similar to the maximum  CR energy gained
at a parallel shock.
If a SNR shock expands across a perpendicular magnetic field configured in the form of a Parker spiral
$uBR$ is the approximate electric potential between the pole and the equator of the SNR in the shock frame.
It is possible to construct special CR trajectories in special magnetic field configurations
for CR to gain higher energies, but these artificial cases are unlikely to be responsible for
the acceleration of CR in large numbers.
Although perpendicular shocks may be rapid CR accelerators,
CR cannot reach PeV energies at SNR shocks propagating into fields
of a few microGauss.

%================ SECTION 4 ===================================================================
\section{Magnetic field amplification}

Apart from a numerical factor of order unity, the `Hillas parameter' $uBR$ 
provides a good estimate of the maximum CR energy where $u$, $B$ and $R$ 
are the characteristic velocity, magnetic field and spatial scalelength respectively.
Hillas (1984) surveyed a large range of putative CR acceleration sites with greatly varying values of
$B$, $u$ and $R$ only to find that none of them stand out as the most likely source of
high energy CR.
At one extreme, 
$B$ and $u$ are very large but $R$ is very small in pulsar magnetospheres.
At the other extreme,  $R$ and $u$ are very large but $B$ is very small
in the lobes of radio galaxies.

Of the three quantities comprising the Hillas parameter, $u$ and $R$ are relatively fixed
by shock kinetics and geometry.
Only $B$ is open to radical increase.
It became apparent both observationally and theoretically in the early 2000s that
the magnetic field can be amplified at high velocity shocks far beyond the unperturbed upstream value.
Vink \& Laming (2003), Berezhko et al (2003) and V\"{o}lk et al (2005) showed that amplified
magnetic fields were needed to explain the thin strands of x-ray synchrotron emission
coincident with the outer shocks of young supernova remnants.
The essence of their analyses was that the thickness of the x-ray strand
is determined by synchrotron losses.
The fact that they are thin indicates that the electron synchrotron loss time
is short and the magnetic field must be large.
The implied fields are of the order of 100s of microGauss downstream of the
shock.
The factor of $\sim 100$ increase in magnetic field correspondingly increases
the Hillas parameter by the same factor and may facilitate CR acceleration to
a few PeV.

%====================== SECTION 5 =============================================================
\section{The non-resonant hybrid (NRH) instability}

Coincidentally at the same time that amplified magnetic field was detected
observationally in SNR, it was shown theoretically that there
exists a plasma instability capable of amplifying magnetic field
by orders of magnitude (Bell 2004).

It had been recognised from the earliest papers on diffusive shock acceleration
that the Alfven instability driven by CR streaming ahead of the shock could generate
circularly polarised Alfven waves
with wavelengths in spatial resonance with the CR Larmor radius.
The spatially resonant Alfven waves strongly scatter the CR 
thereby reducing the CR mean free path and
diffusion coefficient and enabling diffusive shock acceleration.
Because the instability relies upon a spatial resonance between the wavelength
and the CR Larmor radius it seemed unlikely that the instability would grow beyond 
$\delta B/B \sim 1$ since the resonance would then be destroyed.
The instability might reduce the diffusion coefficient to the Bohm value if $\delta B/B \sim 1$,
but would not increase the Hillas parameter and
could not explain acceleration to PeV energies.

The non-linear evolution of the magnetic field under the influence of CR streaming
 was investigated numerically by Lucek \& Bell (2000).
 They coupled a particle simulation of initially streaming CR to a 3D MHD model of the
background plasma.  They found that the magnetic field grew by an order of
magnitude beyond the initial value 
before scattering isotropised the streaming CR and removed the 
force driving instability.
This suggested that CR streaming had the potential to increase the magnetic
field and accelerate CR to higher energies.
It was subsequently shown that the dominant instability is in fact not the
resonant Alfven instability but the previously unrecognised non-resonant hybrid (NRH) instability
(Bell 2004).
Being non-resonant there is less reason for the NRH instability 
to saturate at $\delta B/B \sim 1$.
In some configurations, the linear analysis of its growth is in fact valid
for large amplification of the magnetic field (eg Bell 2005).

The NRH instability can be understood by considering 
CR protons propagating with electric current ${\bf j}_0$ 
parallel to a magnetic field ${\bf B}_0$ in the $z$ direction.
Add a small magnetic field ${\bf B}_\perp$ ($B_\perp \ll B_0$)
perpendicular to  ${\bf B}_0$
 with components in the $(x,y)$ plane.
If the wavelength $2\pi/k$ is much smaller than the CR Larmor radius, the
CR trajectories are undeflected and the current is unperturbed (${\bf j}_1 =0$).
The equation of motion for the background thermal
plasma with density $\rho$ moving with velocity ${\bf u}_\perp$ in the 
perpendicular direction is
$$
\rho \frac {\partial {\bf u}_\perp}{\partial t}=-{\bf j}_0 \times {\bf B}_\perp\ .
 \eqno{(2)}
 $$
The density $\rho$ is constant since the velocity is perpendicular to the
$z$ direction.
The $-{\bf j}_0 \times {\bf B}_\perp$
force operates through the return current $-{\bf j}_0$ carried
by the background plasma to balance the forward CR current.
Alternatively the force can be understood as the momentum-conserving 
reaction against the 
${\bf j}_0 \times {\bf B}_\perp$ force acting on the CR.
The magnetic field is frozen into the background plasma and evolves according
to the ideal MHD equation
$$
\frac {\partial {\bf B}_\perp}{\partial t}=\nabla \times \left ({\bf u}_\perp \times {\bf B}_0 \right )
 \eqno{(3)}
 $$
Equations (2) and (3) can be manipulated to give
$$
\frac {\partial B_x}{\partial t}=
\frac{B_0 j_0}{\rho} \frac {\partial B_y}{\partial z}
\hskip 1 cm
\frac {\partial B_y}{\partial t}=
-\frac{B_0 j_0}{\rho} \frac {\partial B_x}{\partial z}
 \eqno{(4)}
 $$
If ${\bf B}_\perp$ varies harmonically in the $z$ direction parallel to
${\bf j}_0$ and ${\bf B}_0$ in proportion to $\exp (ikz)$ 
there exist purely growing circularly polarized modes
with a growth rate
$$
\gamma = \sqrt { \frac {kB_0 j_0}{\rho} }\ .
 \eqno{(5)}
 $$
No approximations have been made in this derivation apart from the
assumption that the wavelength is much smaller than the CR Larmor radius.
The Larmor radius of a PeV proton in a microGauss magnetic field
is 1 parsec so this is an acceptable assumption except that the
scalelength of structure in the magnetic field must grow to a Larmor radius
for effective CR scattering.
Apart from the small wavelength assumption, the growth equation is valid
for arbitrary non-linearity, ie $B_\perp$ arbitrarily greater than  $B_0$.
This pure mode continues to grow indefinitely with growth rate $\gamma$.
The same applies in some other special configurations (Bell, 2005).
In reality the perturbed magnetic field more likely consists of a mixture of modes which
interfere with each other non-linearly and slow the growth rate,
but numerical simulation shows that the magnetic field can still be amplified by
orders of magnitude (Bell 2004).

The NRH instability differs from the Alfven instability in its polarisation.
Both instabilities produce circularly polarised waves of helical magnetic
field but the polarisation
rotates in opposite directions.
In the Alfven instability, the magnetic field rotates in the same sense
as the spiral CR trajectories to maintain a resonance between the magnetic field ${\bf
B}_0$ and the streaming CR.
In contrast, the NRH polarisation and the CR trajectories are counter-rotating
and non-resonant.
This surprising feature may explain why the NRH instability remained
unrecognised for many years.

%********************************************************* figure 1
\begin{figure*}
\centering
\includegraphics [angle=270,width=15cm]{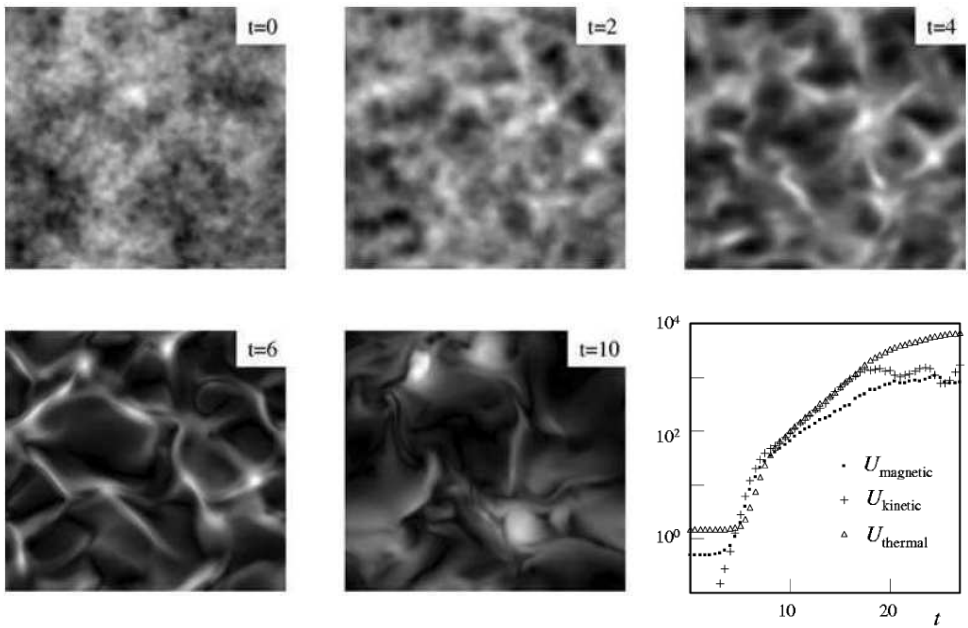}
\caption
{
Plots of the magnitude of a magnetic field driven by CR streaming into the page
(2D slices of a 3D simulation).
The field grows from noise on a small spatial scale at $t=0$.
By $t=10$ the spatial scale has grown to the size of the computational box.
The graph shows how the mean magnetic, kinetic and thermal energy densities increase with time.
This figure is composed from figures 3 and 4 of Bell (2004) where further details can be found.
The units of time are such that $\gamma _{max}=1.26$.
}
%\label{fig:figure1}
\end{figure*}
%********************************************************* figure 1

%========== SECTION 6 =========================================================================
\section{The saturation of the NRH instability}

As shown by the graph of magnetic energy density against time 
in figure 1, the NRH instability can grow non-linearly to amplify
the magnetic field by a large factor.
The instability continues to grow provided two conditions
are satisfied.
Firstly, the instability grows only if its characteristic scalelength $L$
does not exceed the CR Larmor radius,
otherwise CR spiral along the perturbed magnetic field
lines and the ${\bf j}\times {\bf B}$ force averages to zero.
Secondly, the ${\bf j}\times {\bf B}$ force driving
the instability must exceed the magnetic tension 
${\bf B}\times (\nabla \times {\bf
B}/\mu _0)$ which opposes the expansion in the amplitude of spirals in the magnetic field.
To order of magnitude, these two conditions are respectively
$L < p/eB$ and $j> B/L\mu _0$.
Eliminating $L$ between the conditions gives $B^2/\mu _0<Q_{cr}/c$
where $Q_{cr}=jpc/e$ is the energy flux carried by the CR in the 
rest frame of the background plasma.
If $B^2/\mu _0 > Q_{cr}/c$,
either the scalength $L$ must be larger than the CR Larmor radius,
or $L$ must be so short that the magnetic tension overpowers the 
${\bf j}\times {\bf B}$ force.

Note that the resonant Alfven instability is not restricted
in the same way by magnetic tension since the instability
excites Alfven waves which are natural plasma modes and 
propagate without damping provided damping by collisions or wave-wave interactions 
can be neglected.

The NRH instability condition $B^2/\mu _0<Q_{cr}/c$ can be related
to the CR energy density $U_{cr}$.
The upstream CR distribution is nearly at rest in the shock frame,
so $Q_{cr}=uU_{cr}$ in the upstream frame.
If $U_{cr}=\eta \rho u^2$ where $\eta$ is an efficiency factor
then the magnetic energy density is limited to
$B^2/\mu _0  < \eta \rho u^3/c$.

We must be careful about the meaning we attach to
the CR energy density $U_{cr}$, the efficiency $\eta$
and the magnetic energy density $B^2/\mu _0$.
The CR energy spectrum at a SNR shock extends by six orders of magnitude from GeV to PeV.
A GeV proton has a Larmor radius $10^6$ times smaller than that of a PeV proton,
the scaleheight of its precursor ahead of the shock is correspondingly
$10^6$ times smaller, and it drives and responds to magnetic perturbations
which are similarly $10^6$ times smaller.
Hence the system of GeV CR and their accompanying magnetic fluctuations
can be expected to decouple from that of PeV CR and their associated
magnetic fluctuations.
As a consequence, the discussion of $U_{cr}$, $\eta$
and  $B^2/\mu _0$ in the previous paragraph refers to a particular energy range of CR.
Since CR energy densities are spread nearly equally between GeV and PeV,
only a fraction $\sim 1/\log(10^6) \sim 0.07$ of the total CR energy density
contributes to magnetic field amplification on a specified scale.
If $\sim 30\%$ of $\rho u^2$ is given to CR in total,
then $U_{cr}=\eta \rho u^2$ where $\eta \approx 0.3/\log(10^6) \approx 0.03$.  
Bell (2004) and Bell et al (2013) give
more detailed discussions of $\eta$.
See also Zirakshvili \& Ptuskin (2008), although their parameter $\eta _{esc}$
differs in definition by a numerical factor from our $\eta$.
The magnetic energy density associated with a particle energy range of CR
is then $B^2/\mu _0 \approx \eta \rho u^3/c$ with $\eta \approx 0.03$.
This provides an estimate of a magnetic field 
$B=160 n_e^{1/2}u_7^{3/2}\ {\rm \mu G}$ available to scatter CR in a particular energy range.

The magnetic field is amplified on a range of scales 
corresponding to the Larmor radii of CR with energies ranging from GeV to PeV,
in which case the total magnetic energy density immediately upstream of the shock
is better estimated by setting $\eta \approx 0.3$
giving $B=500 n_e^{1/2}u_7^{3/2}\ {\rm \mu G}$.
The field is compressed by a further factor of about 3
when it passes through the shock to give a good match to the
magnetic field inferred from the observed x-ray synchrotron emission at SNR shocks.

%=============== SECTION 7 ====================================================================
\section{NRH scalelengths}

For the NRH instability to explain CR acceleration to PeV it must not only
amplify the magnetic field sufficiently but also produce magnetic structures
on the scale of the CR Larmor radius.
The growth rate is proportional to $\sqrt{k}$ so small scale structure grows
most rapidly.
Fortunately, magnetic field amplification and increase in spatial scalelength
are connected aspects of the non-linear NRH instability
since the field grows by the stretching of field lines.
In its simplest configurations the magnetic field is amplified
by an increase in the radius of loops of magnetic field or in the radius
of a helical field-line.
Equation (3) for a harmonic mode with wavenumber
$k$ can be integrated to show that the radius of spirals in the magnetic
field is $r \sim k^{-1}B_\perp/B_0$, which indicates that the scalelength
grows nonlinearly in proportion to the magnetic field.

Nonlinear numerical simulations presented in the grey-scale plots of figure 1 
(see also Reville et al 2008, Zirakashvili, Ptuskin \& V\"{o}lk 2008, Riquelme \& Spitkovsky 2009, 2010, and Reville \& Bell 2013) 
clearly demonstrate strong growth in scale.
The instability evolves nonlinearly into a series of expanding cells consisting of
walls of high plasma density and strong magnetic field spiralling around
cavities of low density and low magnetic field.
As discussed in the previous section,
the instability grows over a range of wavenumbers bounded by $k>r_g^{-1}$ at long
wavelength and $kB<\mu_{0}j$ at short wavelength.
Saturation occurs when these two wavenumbers coincide 
with both $k\sim r_g^{-1}$ and $kB \sim \mu_{0}j$.
Consequently $kr_g \sim 1$ at saturation and the dominant scalelength 
at saturation is
equal to the CR Larmor radius as required for effective CR scattering.

%============== SECTION 8 ===================================================================

\section{NRH growth times}

As shown above the NRH instability has the capacity to amplify the magnetic field by orders
of magnitude.
We now investigate the conditions necessary for the instability
to grow through many e-foldings in the available time.
The NRH growth time is
$$
\gamma^{-1}=3\times 10^4 (kr_g)^{-1/2}u_7^{-3/2} E_{PeV} B_{\mu G}^{-1} \eta_{0.03} ^{-1/2}
\ {\rm yr}
 \eqno{(6)}
 $$
where $E_{PeV}$ is the energy of CR protons in PeV and $\eta _{0.03}$  is $\eta$ relative to 0.03 
($\eta _{0.03}=1$ if $\eta =0.03$).
Equation (6) shows that the NRH instability does not have time to grow on the scale of 
the Larmor radius ($kr_g\sim 1$) of a PeV proton in young SNR.
Fortunately, the growth rate is proportional to $\sqrt{k}$ so there is time for
the instability to grow initially on a small scale ($kr_g \gg 1$) 
and subsequently expand non-linearly
to match the CR Larmor radius.
The smallest scalelength on which the instability can grow is
set by the tension in the magnetic field.
As discussed above, the ${\bf j}\times {\bf B}$ must dominate the
magnetic force 
${\bf B}\times (\nabla \times {\bf B})/\mu _0  \sim kB^2/\mu _0$ and
this condition is satisfied if $k< \mu _0 j/B$.
The  growth rate is at a maximum $\gamma _{max}=0.5 j\sqrt{\mu _0 /\rho}$
when the wavenumber is $k_{max}=0.5 \mu _0 j/B$ (Bell 2004).
The corresponding fastest growth time and associated scalelength are
$$
\gamma _{max}^{-1}=40 n_e^{-1/2} u_7^{-3} E_{PeV} \eta _{0.03}^{-1} \ {\rm yr}
 \eqno{(7)}
 $$
$$
k_{max}^{-1}=3 \times 10^{12}n_e^{-1} u_7^{-3} E_{PeV} B_{\mu G} \eta _{0.03}^{-1} \ {\rm m}
 \eqno{(8)}
 $$
giving
$$
k_{max} r_g=10^{4} \eta _{0.03} n_e u_7^{3} B_{\mu G}^{-2}
 \eqno{(9)}
 $$
The growth time $\gamma _{max}^{-1}$ of the fastest growing mode is sufficiently
short to allow many e-foldings during the lifetime of a SNR.
The initial scalelength $k_{max}^{-1}$ is much smaller than the CR Larmor radius
but $k_{max}^{-1}$ increases in proportion to the growing magnetic field.
$k_{max} r_g $ $\propto B^{-2}$,
so an increase in magnetic field from a few $\mu G$ to 
$\sim 100 \mu {\rm G}$ increases $k_{max} r_g $ to $\sim 1$
as required for effective CR scattering.

%========== SECTION 9 ======================================================================
\section{$E_{max}$ as determined by NRH growth times}

As seen in the grey-scale plots in figure 1 magnetic field structures 
grow substantially in a few growth times.
The graph of magnetic energy density in figure 1 shows that the magnetic field grows 
substantially in the same time.
In this section we take $\tau_{amp}= 5\gamma_{max}^{-1}$ to be the characteristic time
for magnetic field amplification as observed in young SNR. 
As shown in section 8 of Bell et al (2013) $E_{max}$ 
for a spherical freely expanding blast wave can be calculated by
equating $\tau_{amp}$ to $R/2u$.
Equation (7) can be inverted to derive an equation
for the maximum CR energy:
$$
E_{max}=230 n_e^{1/2} u_7^2 R_{pc} \eta_{0.03}\ {\rm TeV}
 \eqno{(10)}
 $$
An equivalent result was first derived by Zirakashvili \& Ptuskin (2008)
using a more sophisticated model
for CR transport and acceleration.
The physics underlying their calculation of the maximum CR
energy is the same in essence. 

Equation (10) for the maximum CR energy is governed by the NRH growth rate.
If CR were accelerated to a higher energy, 
the upstream CR electric current driving the instability would be reduced for the same CR energy flux
$\eta \rho u^3$
since fewer CR would be needed to carry the CR energy.
Conversely, if CR were accelerated only to a lower energy
the CR electric current would be increased, the growth rate would increase, and the magnetic field
would be amplified sufficiently to accelerate CR to the higher energy.
Hence the CR current and the NRH growth 
are self-regulated to drive the instability through about 
five e-foldings ($\tau_{amp}\approx 5\gamma_{max}^{-1}$)
in the time available.
Note that the magnitude of the magnetic field does not appear in
this derivation of $E_{max}$.

It immediately stands out that $E_{max}$ in equation (10) falls short of the energy of the knee
for parameters appropriate to young SNR.
Our derivation of $E_{max}$ is approximate, for example in the estimation that
$\tau_{amp}=5\gamma_{max}^{-1}$. 
From figure 1 it appears 
that if anything a time greater than $5\gamma_{max}^{-1}$
should be allowed for magnetic field amplification, in which case
$E_{max}$ would be lower and acceleration to PeV less likely.
On the other hand, there are ways in which the value for $E_{max}$ 
given in equation (10) might be an underestimate.
For example,
our derivation treats CR as mono-energetic.
Lower energy CR may be leaking upstream from the shock to add to the electric current driving 
magnetic field amplification.

Equation (10) gives $E_{max}\sim 100-200$TeV 
for the historical SNR since their velocities are 
$\sim 5000 {\rm km\ s}^{-1}$ or less,
which is a full factor of ten lower than the CR energy at the knee.
The CR energies reached in the historical SNR and the implications
for Galactic CR origins are discussed below in section 14.

%============ SECTION 10 =====================================================================
\section{CR escape upstream is essential}

Here we emphasise that CR acceleration only occurs if 
CR accelerated to the highest energy escape upstream of the shock
in contrast to CR accelerated to lower energies which are advected away downstream of the shock.

Consider a point ahead of the approaching shock.
CR streaming away from the shock pass freely through this point until the
instability has grown sufficiently to scatter CR and confine them to the
shock precursor.
In accord with the discussion above, this occurs when $\int \gamma_{max} dt \sim 5$
where $\gamma_{max}=0.5 j\sqrt {\mu _0 /\rho}$.
Consequently, confinement sets in when the time integral of CR charge per unit area
passing through this point reaches
$$
q_{cr}=\int j dt \sim 10 \sqrt {\rho / \mu _0}
 \eqno{(11)}
 $$
The condition for confinement is that a charge $q_{cr}$ per unit area of shock 
must escape freely upstream
before CR following after it can be confined.
Hence CR escape upstream is an essential feature of CR acceleration by shocks.

%========== SECTION 11 ===========================================================================

\section{The shock velocity at which the NRH instability switches off}

The NRH instability operates (i) only on scales comparable or smaller than the Larmor radius 
of the CR with energy $E$ (in eV) driving the instability
(ie $k>cB/E$), 
and 
(ii) only if the ${\bf j}\times {\bf B}$ force driving magnetic field amplification is stronger
than the magnetic force ${\bf B}\times (\nabla \times {\bf B})/\mu _0$
(ie $k<\mu_0 j/B$).
Consequently 
there is no range of wavelengths at which the NRH instability
is active if $\mu_0 j/B<cB/E$, (ie $j < cB^2/\mu _0 E$).
Since the CR energy flux is $jE=\eta \rho u^3$,
the condition for magnetic field amplification by the NRH instability is
$u > (c B^2/\eta \rho \mu _0)^{1/3}$,
which is equivalent to
$$
u> 350 B_{\mu G}^{2/3} \eta _{0.03}^{-1/3} n_e ^{-1/3}\ {\rm km\ s}^{-1}
 \eqno{(12)}
 $$
This limit implies that the NRH instability is active in the historical SNR 
which are expanding at $\sim 5000 {\rm km\ s}^{-1} $,
but that it is unavailable for magnetic field amplification 
during much of the Sedov phase.
When the expansion drops below about  $1000 {\rm km\ s}^{-1} $ the
${\bf j}\times {\bf B}$ force is too weak to overcome
the magnetic force ${\bf B}\times (\nabla \times {\bf B})/\mu _0$
for magnetic fields of a few $\mu {\rm G}$.
This condition (equation 12) is equivalent to requiring that the saturation magnetic field
derived in section 6 must be greater than the initial
magnetic field in the ambient medium.
It does not apply to the Alfven instability 
where the magnetic force is part of the
natural oscillation of the Alfven waves excited by the instability
as mentioned in section 6.

%=========== SECTION 12 ==========================================================================

\section{The Alfven instability}

As shown above, the NRH instability at the outer shocks of SNR is
only excited in the early stages of evolution when the expansion
velocity is comparable with or greater than $\sim 1000 {\rm km\ s}^{-1} $
for an ambient magnetic field of a few $\mu{\rm G}$.
Below this velocity, the resonant Alfven instability is dominant.

From equations (6) and (7) in Bell (2004) the dispersion relation including both the NRH and
Alfven instabilities for a mono-energetic CR distribution drifting diffusively at the shock velocity $u$ is
$$
\omega ^2
= k^2 v_A^2
- \frac{kjB}{\rho} \times \hskip 6 cm
$$
$$\left \{
1 - \frac {3}{2k^2 r_g^2}+\frac{3}{4}\frac{1-k^2 r_g^2}{k^3 r_g^3}
\left [ i\pi + \log \left( \frac{kr_g+1}{kr_g-1}\right )\right ]
\right \}
 \eqno{(13)}
 $$
where $r_g$ is the CR Larmor radius and the imaginary term on the right hand side
is cancelled when $ k r_g <1 $ by the imaginary part of the logarithmic term.
The combination of the quantities $kjB$ shows the importance of the 
relative directions of CR current, the magnetic field and the polarisation of the wave.
Changing the sign of $j$ while keeping $k$ and $B$ unchanged
causes an unstable wave to become stable and vice versa under most circumstances.
See Bell (2004) for more detail.

In the limit of large $j$ 
appropriate to young SNR 
the NRH instability grows most strongly at 
wavelengths shorter than the CR Larmor radius ($ k r_g \gg 1 $) where
the dispersion relation reduces to 
$$
\omega ^2
= k^2 v_A^2 - \frac{kjB}{\rho}
 \eqno{(14)}
 $$
and the term $k^2 v_A^2$  sets the short wavelength limit beyond which the 
magnetic tension is stabilising.

In the limit of small $j$ ($jB/\rho \ll v_A^2/r_g$)
appropriate to old SNR 
and growth of the Alfven instability
the dispersion relation reduces to
$$
\omega ^2
= k^2 v_A^2
+ \frac{3\pi i}{4} \frac{kjB}{\rho}
\frac{k^2 r_g^2 -1}{k^3 r_g^3}
 \eqno{(15)}
 $$
The maximum instability growth rate of the Alfven instability is
$$
\gamma _{A,max}=\frac{\pi}{4\sqrt{3}} j \sqrt {\frac{\mu _0}{\rho}}
 \eqno{(16)}
 $$
which occurs when the wavelength of the Alfven wave is $3.6 r_g$ ($kr_g=\sqrt{3}$).

The growth of the Alfven instability is governed by the imaginary term $(i\pi)$ in the
square brackets in equation (13) in contrast to the growth of the NRH
which is determined by the first term since fastest NRH growth occurs when
$kr_g>>1$.
The Alfven instability is driven by ${\bf j}_1 \times {\bf B}_0$
where ${\bf j}_1$ is the perturbed CR current and ${\bf B}_0$ is the unperturbed
magnetic field, whereas the NRH instability is driven by ${\bf j}_0\times {\bf B}_1$.
The two instabilities also differ in their polarisation and dominant wavenumber:
$kr_g \approx 1$ for the Alfven instability compared with 
$kr_g > 1$ for the fastest growing NRH wavelength.
Both instabilities have nearly the same maximum growth rate,
but each dominates under different conditions.
The crossover from the Alfven to the NRH instability occurs when the CR current $j$
becomes large enough for $kjB/\rho$ to exceed $k^2v_A^2$ on the Larmor scale ($kr_g \sim 1$)
at which point the
${\bf j} \times {\bf B}$ force overpowers
the tension in the magnetic field on the scale of a CR Larmor radius.

The similar values of the Alfven and NRH maximum growth rates,
as pointed out by Zirakashvili \& Ptuskin (2008), 
allows the analysis of section 9 to be applied at low shock velocities relevant to older SNR
as well as to high velocity young SNR shocks at which the NRH instability is active.
Equation (10) is therefore applicable to both young SNR
and SNR throughout the Sedov phase.

Equation (10) finally becomes inapplicable when the shock 
velocity drops to a few times the Alfven speed,
$v_A=2 B_{\mu G} n_e^{-1/2}{\rm km\ s}^{-1}$
since the Alfven instability is driven by the drift of CR relative to the motion
of the the driven resonant Alfven wave.
However, the shock may 
have dissipated by that time if
the shock velocity has dropped below the ion acoustic sound speed, 
$c_s=13 T_{eV}^{1/2} {\rm km \ s}^{-1}$.

%==== SECTION 13 =================================================================================

\section{Spectrum of escaping CR}

Since the maximum growth rates of both the NRH and the Alfven instabilities
are the same to within a numerical factor close to one,
equation (10) for the energy $E_{max}$ of escaping CR can be applied throughout the
lifetime of a SNR.
By definition of $\eta$, the energy flux of CR
escaping with energy $E_{max}$ is $\eta \rho u^3$
and the total energy spectrum of CR injected into the Galaxy
by a SNR can be derived.

The number $N(E)$ of CR escaping with energy greater than $E$ is the number of CR escaping before the
 SNR radius expands to a radius R at which $E_{max}=E$.
$$
N(E)= \int _0^R 4 \pi R^2 \frac{\eta \rho u^2}{E} dR
 \eqno{(17)}
 $$
The corresponding differential energy spectrum $n(E)=-dN/dE$ is
$$
n(E)=-4 \pi R^2 \frac{\eta \rho u^2}{E} \frac{dR}{dE}
 \eqno{(18)}
 $$
From equation (10), $E\propto R u^2$ for a uniform ambient density. 
In the self-similar Sedov phase $u^2 \propto R^{-3}$,
so the spectrum of CR escaping into the interstellar medium
in the Sedov phase is
$$
n(E) \propto E^{-2}
 \eqno{(19)}
 $$
which is the same as the spectrum for test particle CR 
accelerated at a strong shock.
This is not surprising since
the same $E^{-2}$ spectrum of escaping CR can arise under
a range of self-similar assumptions regarding
CR escape during the Sedov phase
(Caprioli et al (2010), Drury (2011), Ohira et al (2010)).
It is sufficient (from equation 18) that the dependence of the CR escape energy 
on shock radius should be a power law 
and that the efficiency $\eta$ should be constant.

The energy of a Sedov blast wave of radius $R$ 
expanding into a medium with $\gamma=5/3$ and density $\rho$
at velocity $u$ is approximately $3\rho R^3 u^2$
with the consequence from equation (10) that
the energy of CR escaping at a particular stage of a Sedov expansion is
$$
E=5 M_A^{4/3}  \epsilon _{44}^{1/3} B_{\mu G}^{4/3} n_e^{-1/2} {\rm GeV}
 \eqno{(20)}
 $$
where the expansion velocity is given in terms of the Alfven Mach number $M_A$ of the shock, 
$\epsilon _{44}$ is the blast
wave energy in $10^{44}{\rm J}$ and $B_{\mu G}$ is the ambient magnetic field
in $\mu{\rm G}$.
The shock dissipates into the interstellar medium as $M_A \rightarrow 1$
at which time CR are no longer confined to the remnant since the magnetic
fluctations that scatter them move at the Alfven speed.
Assuming that CR acceleration weakens when $M_A\sim 3$ at a shock propagating
into a field of $5 \mu {\rm G}$, the energy spectrum of escaping CR
terminates at a lower energy of $E_{min} \sim 200 {\rm GeV}$.
At this point CR previously confined to the interior of the SNR are released
into the interstellar medium.

Galactic CR accelerated by SNR 
can be divided into two populations,
(i) CR above $E_{min} \sim 200 {\rm GeV}$ that escape upstream during SNR expansion
(ii) CR at energies below $E_{min}$, but also extending into the TeV range, that are
advected into the interior of the SNR and released into the ISM at the end
of SNR expansion when the shock dissipates.

For the different reasons explained above,
both populations have the same energy spectrum $\propto E^{-2}$ at source,
although in reality
additional factors might be expected to slightly steepen or flatten
each spectrum, and propagation losses steepen the entire spectrum to the approximately
$ E^{-2.7}$ power law arriving at the Earth.
It is unlikely that the spectra of the two populations should join smoothly at $E_{min}$ without
at least a slight bend in the spectrum or some noticeable feature.
The most obvious candidate for a spectral signature of a join in the spectrum
is that seen  by PAMELA at about 200GeV (Adriani et al 2011). 
The observational
status of the detailed spectral feature is uncertain in the light of 
recent AMS observations (Ting 2013), 
but a change in spectral index near this energy is supported by other measurements (Ahn et al 2010).
Also our theoretical estimate of 200GeV as the exact value of $E_{min}$ is subject
to considerable uncertainty,
and energy-dependent CR propagation might be an alternative explanation for a
change in spectral index (Blasi et al 2012, Tomassetti 2012).

The total energy in escaping CR can be derived from the current
$j_{cr}$ of escaping CR as discussed above (equations 17 \& 18).
Only the highest energy CR being accelerated escape at any one time.
They carry only $\sim 10 \%$ 
of the energy content of the CR carried into the interior of the SNR.
CR carried into the interior cool adiabatically as
the SNR expands.
Their energy is recouped as their pressure drives the shock and accelerates later generations of CR.
Integration over the whole lifetime of the SNR 
shows overall that the acceleration and release of CR into
the interstellar medium can be a highly efficient process (Bell et al 2013).

%==============  SECTION 14 =======================================================================

\section{Can SNR accelerate CR to the knee?}

The expression,
$
E_{max}=230 n_e^{1/2} u_7^2 R_{pc} {\rm TeV}
$, 
derived in section 9, equation (10), for the maximum CR energy can be applied to the
historical SNR usually regarded as the most likely source of PeV CR in
the Galaxy.
Similar results are obtained by Zirakashvili \& Ptuskin (2008).
The controlling parameters, $n_e$, $u_7$ and $R_{pc}$, are often uncertain, particularly the ambient density
$n_e$, but also the radius $R_{pc}$ since the distance to the SNR is 
not always well determined.
The predicted maximum CR energies in TeV are
\newline
Cas A: 160 ($n_e=2$, $R_{pc}=2$ \& $u_7=0.5$)
\newline
SN1006: 100 ($n_e=0.3$, $R_{pc}=9$ \& $u_7=0.3$)
\newline
Tycho: 100 ($n_e=0.2$, $R_{pc}=4$ \& $u_7=0.5$)
\newline
Kepler: 70 ($n_e=1$, $R_{pc}=2$ \& $u_7=0.4$)
\newline
(Raymond et al 2007, Vink 2008, Williams et al 2013).
The maximum CR energy is low in each case because
the expansion is already heavily decelerated.
If we replace the current expansion velocity with the mean expansion velocity
$\bar{u}_7=R_{pc}/t_{100}$ where $t_{100}$ is the age in hundreds of years,
such that equation (10) becomes 
$
E_{max}=230 n_e^{1/2} R_{pc}^3 t_{100}^{-2} {\rm TeV}
$,
then the estimated maximum CR energies are larger:
\newline
Cas A: 240 ($n_e=2$, $R_{pc}=2$ \& $t_{100}=3.3$)
\newline
SN1006: 900 ($n_e=0.3$, $R_{pc}=9$ \& $t_{100}=10$)
\newline
Tycho: 540 ($n_e=0.5$, $R_{pc}=4$ \& $t_{100}=4.4$)
\newline
Kepler: 120 ($n_e=1$, $R_{pc}=2$ \& $t_{100}=4$)
\newline
The mean expansion velocity predicts a higher CR energy in each case,
approaching 1PeV for SN1006 because of its high average expansion velocity.
However, the predicted CR energy depends on the third power of the radius and hence distance to the SNR.  
The derived distances to Tycho and Kepler vary by a factor of two between
the largest and the smallest, so the uncertainties are very large in
the above CR energies. 
Moreover equation (10) might easily be
in error by a factor of 2 or 3
because of the assumptions entering its derivation as discussed in section 9.
Most of the assumptions made in the derivation of equation (10) were generous in terms of
trying to make $E_{max}$ larger rather than smaller so as to reach the knee.
For example it is easy to see how the efficiency parameter $\eta$
might be smaller than $0.03$ but not much larger than this
(but see Schure \& Bell 2013b).
Nevertheless,
the CR energies predicted for present expansion velocities
fall substantially short of the energy of the knee,
so it appears likely that none of the historical SNR
are at present accelerating CR to the knee.

The dependence of $E_{max}$ on the square of the shock velocity implies that the
highest CR energies are more probably reached in young high velocity SNR.
The youngest known Galactic SNR, G1.9+0.3 aged 110 years, is expanding at
$18,000{\rm km\ s}^{-1}$ 
with a radius of 2.2pc (Borkowski et al 2013). 
Reynolds et al (2008) give an ambient density of $0.04{\rm cm}^{-3}$
predicting a maximum CR energy $E_{max}=330{\rm TeV}$ from equation (10).
If the SNR were expanding at the same velocity into a higher density circumstellar medium,
$1\ {\rm cm}^{-3}$ for example, it could be accelerating CR to the knee.

However uncertain the above estimates for $E_{max}$,
the strong dependence on expansion velocity points to the ideal PeV accelerator 
as a very young SNR expanding at high velocity
into a dense wind environment as previously suggested by 
V\"{o}lk \& Biermann (1988) and Bell \& Lucek (2001).
Ptuskin et al (2010) survey different types of supernovae and find that
type IIb SN may accelerate CR to a particularly high energy,
and may accelerate iron to $5\times 10^{18}{\rm eV}$.
SN1987A may be a relatively nearby and well-diagnosed 
example of a high energy accelerator.
Constant expansion for time $t_{yr}$ in years at its
initial expansion velocity of $35,000{\rm km\ s}^{-1}$
predicts
$E_{max}=100 t_{yr} n_e^{1/2} {\rm TeV}$.
Expansion into a high density circumstellar medium over a decade or more
should accelerate CR to the knee quite easily, and possibly to higher energies. 
SN1987a has passed through a number of different phases since
it exploded into a highly structured medium
including rings of radius 0.2pc and with a range of densities 
$\sim (1-30) \times 10^3 {\rm cm}^{-3}$ (Mattila et al 2010).
The ejecta velocity slowed to $7,000{\rm km s}^{-1}$ on encountering the dense ring
(Larsson et al 2011).
Substituting $u_7=0.7$, $n_e=10^4$ \& $R_{pc}=0.2$ in equation (10)
predicts an estimated maximum CR energy at that stage of $E_{max}=2{\rm PeV}$.
The ring has a mass of $\sim 0.06$ solar masses (Mattila et al 2010), so
the energy processed through the shock during the encounter with the ring
is $\sim 3\times 10^{42}{\rm J}$ which would contribute substantially 
to the Galactic CR energy budget if a similar SNR occurred in the Galaxy.

It appears that very young SNR are the most likely
sites for acceleration to the knee.
Schure \& Bell (2013a) have analysed the evolution of the CR spectrum in
the pre-Sedov phase for SNR expanding into a pre-ejected circumstellar wind.
They find that Cas A probably accelerated CR to PeV energies
at an earlier stage of its expansion.
If acceleration to the knee only happens where SNR expand into a dense wind
then the CR composition can be expected to be heavier at energies approaching
the knee.
This would be consistent with the observation that the Galactic helium spectrum
is flatter than the proton spectrum
and that helium nuclei are more numerous than protons at the knee (Kampert
2007).

%================ SECTION 15 ===============================================================

\section{Conclusion}

Cosmic ray physics is now solidly part of mainstream astrophysics,
particularly as a result of x-ray and $\gamma $-ray observations
of in situ CR acceleration.
Theoretical models of Galactic
CR acceleration to PeV energies can be constructed with the tentative conclusion
that CR are accelerated to the knee in very young SNR and
that probably none of the known Galactic SNR are
able to accelerate protons beyond a PeV at the present time. 
Measurements of CR composition and precise determinations
of the detailed energy spectra of different species
increasingly provide powerful tests of theory.
Observations with the CTA Cherenkov telescope (Hinton \& Hofmann 2009, Aharonian 2013)
should decisively indicate whether our 
current models are on the right lines.

%=====================================================================================

\section{Acknowledgements}

I thank Brian Reville, Klara Schure and Gwenael Giacinti
for many insightful discussions on cosmic ray and related physics.
The research leading to this review has received funding
from the European Research Council under the European
Community's Seventh Framework Programme (FP7/2007-
2013)/ERC grant agreement no. 247039 and from grant number ST/H001948/1
made by the UK Science Technology and Facilities Council. 

%=====================================================================================

\newpage

\section{References}

Adriani O et al, 2011, Science 332, 69

\noindent 
Aharonian F, 2013, Astropart Phys 43, 71

\noindent
Aharonian F et al, 2007, A\&A 464, 235

\noindent
Aharonian F et al, 2011, A\&A 531, C1

\noindent
Ahn HS et al, 2010, ApJLett 714, L89

\noindent
Axford WI Leer E Skadron G, 1977, Proc 15th ICRC (Plovdiv)
11, 132

\noindent
Bell AR, 1978, MNRAS 182, 147

\noindent
Bell AR, 2004, MNRAS 353, 550

\noindent
Bell AR, 2005, MNRAS 358, 181

\noindent
Bell AR, 2013, Astropart Phys 43, 56

\noindent
Bell AR \& Lucek SG 2001 MNRAS 321 433

\noindent
Bell AR, Schure KM, Reville B, Giacinti G, 2013, MNRAS 431, 415

\noindent
Berezhko EG, Ksenofontov LT \& V\"{o}lk HJ, 2003, A\&A 412, L11

\noindent
Blandford RD \& Eichler D, 1987, Phys Rep 154, 1

\noindent
Blandford RD \& Ostriker JP, 1978, ApJ 221, L29

\noindent
Blasi P \& Amato E, 2012, JCAP01(2012)011 

\noindent
Blasi P, Amato E \& Serpico PD, 2012, Phys Rev Lett 109, 061101

\noindent
Borkowski KJ, Reynolds SP, Hwang U, Green DA, Petre R, Krishnamurthy K \& Willett R,
2013, ApJL 771, 9

\noindent
Bykov AM, Osipov SM \& Ellison DC, 2011, MNRAS 410, 39

\noindent
Caprioli D, Blasi P \& Amato E, 2010, Astropart Phys 33, 160

\noindent
Drury LO'C 1983, Rep Prog Phys 46, 973

\noindent
Drury LO'C., 2011, MNRAS 415, 1807

\noindent
Drury LO'C \& Falle SAEG, 1986, MNRAS 222, 353

\noindent
Ellison DC, Baring MG \& Jones FC, 1995, ApJ 453, 873

\noindent
Fermi E, 1949, Phys Rev 75, 1169

\noindent
Giacalone J \& Jokipii JR, 2007, ApJL 663, L41

\noindent
Guo F, Li S, Li H, Giacalone J, Jokipii JR \&  Li D, 2012,
ApJ 747, 98

\noindent
Hillas AM, 1984, ARA\&A 22, 425

\noindent
Hinton JA \& Hofmann W, 2009, ARA\&A 47, 523

\noindent
Jokipii JR, 1987, ApJ, 313, 842

\noindent
Jones FC \& Ellison DC, 1991, Space Sci Rev 58, 259

\noindent
Kampert K-H, 2007,  Nucl Phys B Proc Supp 165, 294

\noindent
Krymsky GF, 1977, Sov Phys Dokl 22, 327

\noindent
Krymsky GF et al, 1979, Proc 16th ICRC (Kyoto) 2, 39

\noindent
Kulsrud R \& Pearce WP, 1969, ApJ 156, 445

\noindent
Lagage O \& Cesarsky CJ, 1983a, A\&A 118, 223

\noindent
Lagage O \& Cesarsky CJ, 1983b, ApJ 125, 249

\noindent
Larsson J et al, 2011, Nature 474, 484

\noindent
Lerche I, 1967, ApJ 147, 689

\noindent
Lucek SG \& Bell AR, 2000, MNRAS 314, 65

\noindent
Malkov MA \& Drury LO'C, 2001, Rep Prog Phys 64, 429

\noindent
Mattila S et al, 2010, ApJ 717, 1140

\noindent
Meli A, Biermann PL, 2006, A\&A 454, 687

\noindent
Ohira Y, Mirase K \& Yamazaki R, 2010, A\&A 513, A17

\noindent
Ptuskin VS, Zirakashvili VN \& Seo E-S, 2010, ApJ 718, 32

\noindent
Raymond JC et al, 2007, ApJ 659, 1257

\noindent
Reville B \& Bell AR, 2012, MNRAS 419, 2433

\noindent
Reville B \& Bell AR, 2013, MNRAS 430, 2873

\noindent
Reville B, O'Sullivan, S Duffy P \& Kirk JG, 2008, MNRAS 386, 509

\noindent
Reynolds SP, Borkowski KJ, Green DA, Hwang U, Harrus I \& Petre R, 2008, ApJL 680, L41

\noindent
Riquelme M \& Spitkovsky A, 2009, ApJ 694, 626

\noindent
Riquelme M \& Spitkovsky A, 2010, ApJ 717, 1054

\noindent
Schure KM \& Bell AR, 2011, MNRAS 418, 782

\noindent
Schure KM \& Bell AR, 2013a, MNRAS 435, 1174

\noindent
Schure KM \& Bell AR, 2013b, MNRAS in press, arXiv:1310.7027 

\noindent
Skilling 1975a, MNRAS 172, 55

\noindent
Skilling 1975b, MNRAS 173, 245

\noindent
Skilling 1975c, MNRAS 173, 255

\noindent
Stage MD, Allen GE, Houck JC \& Davis JE, 2006, Nature Physics 2, 614

\noindent
Ting S, 2013, '{\it The AMS spectrometer on the Internation
Space Station}', Highlight Talk, ICRC 2013

\noindent
Tomassetti N, 2012, ApJL 752, L13

\noindent
Uchiyama Y et al, 2007, Nature 449, 576

\noindent
Vink J, 2008, ApJ 689, 231

\noindent
Vink J \& Laming JM, 2003, ApJ 584, 758

\noindent
V\"{o}lk HJ, Berezhko EG, Ksenofontov LT, 2005, A\&A 433, 229

\noindent
V\"{o}lk HJ \& Biermann P, 1988 ApJL 333, L65

\noindent 
Williams BJ et al, 2013, ApJ 770, 129

\noindent
Wentzel DG, 1974, ARA\&A 12, 71

\noindent
Zirakashvili VN \& Ptuskin VS, 2008, ApJ 678, 939

\noindent
Zirakashvili VN, Ptuskin VS \& V\"{o}lk HJ, 2008, ApJ 678, 255

\end{document}